\documentclass[12pt,preprint]{aastex}

\shorttitle{Polar Field Reversal}
\shortauthors{Svalgaard & Kamide}

\begin{document}

\title{Asymmetric Solar Polar Field Reversals}

\author{Leif Svalgaard}
\affil{W.W. Hansen Experimental Physics Laboratory, Stanford University, Stanford, CA 94305}
\email{leif@leif.org}
\and
\author{Yohsuke Kamide}
\affil{Rikubetsu Space and Earth Science Museum, Rikubetsu, Hokaido, Japan;}
\affil{Nagoya University, Nagoya, Aichi, Japan}

\begin{abstract}
The solar polar fields reverse because magnetic flux from decaying sunspots moves towards the poles, with a preponderance of flux from the trailing spots. Let us assume that there is a strong asymmetry in the sense that all activity is in the Northern Hemisphere, then that excess flux will move to the North Pole and reverse that pole, while nothing happens in the South. If later on, there is a lot of activity in the South, then that flux will help reverse the South Pole. In this way, we get two humps in solar activity and a corresponding difference in time of reversals. Such difference was first noted by \cite{bab59} from the very first observation of polar field reversal just after the maximum of the strongly asymmetric solar cycle 19. At that time, the Southern Hemisphere was most active before sunspot maximum and the South Pole duly reversed first, followed by the Northern Hemisphere more than a year later, when that hemisphere was most active. Solar cycles since then have had the opposite asymmetry, with the Northern Hemisphere being most active early in the cycle. Polar field reversals for these cycles have as expected happened first in the North. This is especially noteworthy for the present solar cycle 24. We suggest that the association of two peaks of solar activity when separated by hemispheres with correspondingly different times of polar field reversals is a general feature of the cycle.
\end{abstract}


\section{Introduction}

In their epoch-making paper \citep{bab55} the Babcocks summarize their observations of weaker magnetic fields on the sun made possible by H. W. Babcock's invention of the solar magnetograph \citep{bab53}. Their findings have stood the test of time and include the following features: A {\it General Magnetic Field}, usually limited to heliographic latitudes greater than 55\degr, but with occasional extensions towards the equator. {\it Bipolar Magnetic Regions (BMRs)} in lower latitudes appearing as contiguous areas of opposite magnetic polarity obeying Hale's polarity laws, containing Ca II plages and, especially when the regions are young, occasionally sunspots. Filaments occur at the boundaries of regions or, alternatively, divide regions into parts of opposite polarity. As the regions age, they weaken and expand until lost in the background of irregular weak fields. And, occasionally, extended {\it Unipolar Magnetic Regions (UMRs)} of only one polarity, which can have a duration of many months as sources of recurrent geomagnetic storms. As noted, these observations provided objective evidence for the until then only inferential hypothesis that magnetic fields are fundamental to sunspots, plages, prominences, chromospheric structure, coronal and radio emissions, and ejection of neutral but ionized matter. 

In 1959 H. D. Babcock reported \citep{bab59} that the General Field had reversed polarity: ``the south polar field reversed its sign between March and July, 1957. The sign of the north polar field, however, remained positive until November, 1958, when it rather abruptly became negative. For more than a year, the unexpected peculiarity was presented of two poles with the same sign". With the passing of time we find that such behavior is quite common and may have a simple explanation.

\section{Polar Field Reversal}

In his celebrated 1961 paper H. W. Babcock \citep{bab61} lists the reversal of the General Field as the first observation that must be explained by a theory of the solar cycle. Today we would rather use the narrower concept of Polar Fields rather than that of \textit{a} General Field, as evidently the polar fields reverse at different times. Nevertheless, it is clear that the polar fields play a crucial role in the solar cycle, likely causative or at least symptomatic. In Babcock's phenomenological model ``preceding parts of BMRs expand toward the equator as they age, to be neutralized by merging; following parts expand or migrate polewards so that their lines of force neutralize and then replace the initial dipolar field. The result, after sunspot maximum, is a main dipolar field of reversed polarity".

\begin{figure}
\epsscale{1}
\plotone{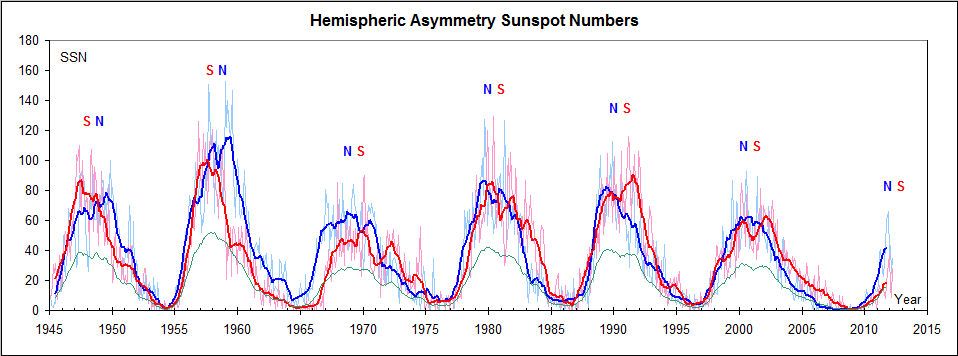} 
\caption{
Sunspot Numbers since 1945 separated by hemisphere (blue - north; red - south). The thin curves show monthly means while the thick curves show the smoothed means. Each cycle is annotated with an estimate of which hemispheres were the most active before and after solar maximum. For reference, the thin (green) line at the bottom shows the full-disk smoothed average, scaled down by a factor of four. Data before 1992 are from \cite{tem06} and thereafter from SIDC (\protect\url{http://www.sidc.be/sunspot-data/dailyssn.php})
\label{SSN}}
\end{figure}

Let us assume that there is a strong asymmetry in the sense that all activity is in the Northern Hemisphere, then that excess trailing flux will move to the North Pole and reverse that pole, while nothing happens in the South. If later on, there is a lot of activity in the South, then that flux will help reverse the South Pole. In this way, we get two humps in solar activity, one in each hemisphere, and a corresponding difference in time of reversals. As noted above, such difference was first observed by \cite{bab59} from the very first observation of polar field reversal just after the maximum of the strongly asymmetric solar cycle 19. At that time, the Southern Hemisphere was most active before sunspot maximum and the South Pole duly reversed first, followed by the Northern Hemisphere more than a year later, when that hemisphere became most active. Solar cycles since then have had the opposite asymmetry, with the Northern Hemisphere being most active early in the cycle, Figure \ref{SSN}. Polar field reversals for these cycles have as expected happened first in the North, as documented in the following section.

\section{Observations}

The net flux of the polar fields comes from a number of strong flux concentrations of vertical kilo-gauss elements with the overwhelmingly same polarity \citep{sva78,shi12}. Small elements and horizontal fields generally cancel out when averaged over the polar cap. During the course of a year, the solar rotation axis tips away from (N pole, March 7) and towards (N pole, September 9) the observer by 7.15 degrees. This, combined with a strong concentration of the flux near the pole and projection effects stemming from the line-of-sight field measurements, causes the observed polar fields to vary by a factor of up to two through the year \citep{sva78,bab55}. Because of the large aperture of the Wilcox Solar Observatory (WSO) magnetograph, the net magnetic flux over the aperture will be observed to be zero (the ``apparent" reversal) well before the last of the old flux has disappeared as opposite polarity flux moving up from lower latitudes begins to fill the equatorward portions of the aperture. We are interested more in the ``true" reversal when the last vestiges of the previous cycle's flux disappear and so shall use the higher-resolution magnetograms from Mount Wilson Observatory (MWO) to determine the time of polar field reversals.

\begin{figure}
\epsscale{1}
\plotone{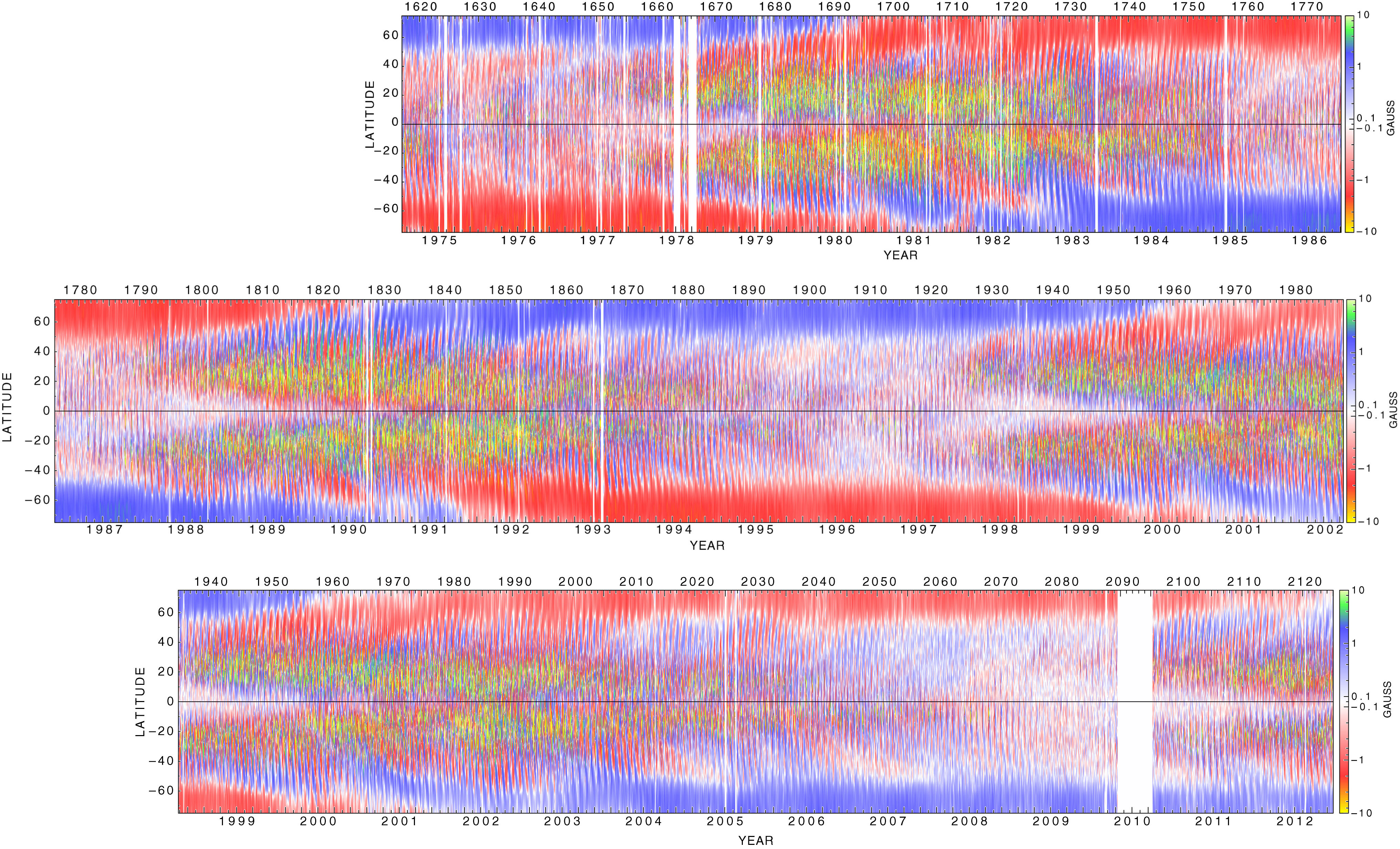} 
\caption{
MWO Supersynoptic maps of the polar reversals at the solar maxima in 1980, 1990, 2001, and (upcoming after) 2012. A supersynoptic map consists of a large number of time-compressed and time-reversed ordinary synoptic maps stacked sideways in time (Years at the bottom of each panel and Carrington Rotations at the top). Positive (away from the surface) polarity is in blue (and green), while negative polarity (towards the surface) is in red (and yellow). Courtesy Roger Ulrich/MWO \citep{ulr02} (\protect\url{http://obs.astro.ucla.edu/torsional.html}).
\label{MWO}}
\end{figure}

The MWO supersynoptic maps, Figure \ref{MWO}, show how the trailing polarity moves polewards, canceling out the old polarity as it goes. For the four recent cycles depicted, the North pole clearly reverses first in every cycle. The vertical `stripes' show another characteristic of the reversal: that poleward migration has a strong longitudinal component, taking place in the same longitude interval over extended periods of time, rather than on a broad `front' advancing in latitude.

\begin{figure}
\epsscale{1}
\plotone{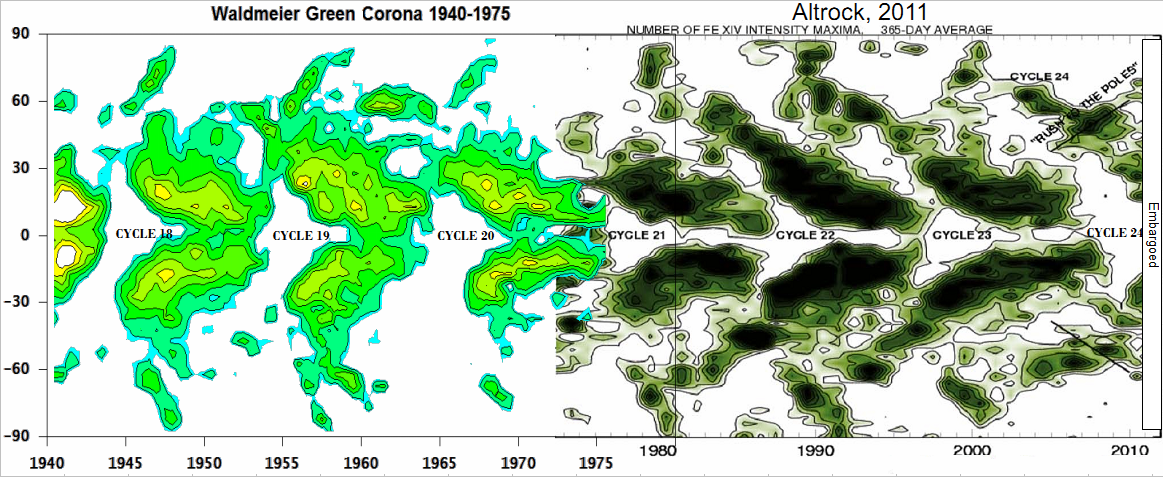}  
\caption{
Contour maps of the distribution in latitude of Green Corona maxima derived (left) from observations by \cite{wal78} and (right) from observations by \cite{alt11}. The `Rush to the Poles' \citep{alt03} delineates in parallel the poleward migration of the polar cap boundary, an indicator of the reversal process.	
\label{Rush}}
\end{figure}

In Cycle 20, peaking about 1969, the pre-maximum activity was strongly concentrated in the North, so the North polar field should reverse first. Unfortunately, the polar fields were weak and could not be clearly observed above the magnetograph noise making it difficult to determine the precise timing of the reversals \citep{how72}. It is well-known that prominences, filaments, and microwave and green corona emissions can give further information about the polar field extent, in particular, the `Rush to the Poles' (RTP) phenomenon \citep{alt03,gop03} that signals the cessation of high-latitude activity is a marker for the time of polarity reversal. Figure \ref{Rush} shows that the RTP for cycle 20 happened in the North well before the South, although matters are complicated by the presence of a secondary prominence zone \citep{wal73}.

The asymmetry is particularly clear for the current cycle 24, where the North polar flux and the northern pole coronal hole have already practically disappeared (see \cite{kir09} for the relationship between coronall holes and polar flux), while the South polar flux has only decreased slightly, as has been noted by several authors, e.g.  \cite{hoe12}, \cite{shi12}, \cite{alt12}, and \cite{gop12}. In analogy with cycle 19, we might expect the South polar fields to reverse, perhaps abruptly, as activity eventually picks up in the Southern Hemisphere (as it already has at the time of writing). 

\section{Conclusion}

The two hemispheres are only weakly coupled and develop rather independently, especially when it comes to polar field reversals. One can look at the solar cycle as a continuous conversion of poloidal field to toroidal field and back to poloidal field. While the generation of toroidal field is probably a rather deterministic and orderly process, the generation of poloidal field seems to be a much more random process, as only a very small fraction (1 to 2\%) of the toroidal field is converted to polar fields by diffusion and/or circulation. E.g.  \cite{cho07} argue that the Babcock{\sbond}Leighton mechanism, in which the poloidal field is produced from the decay of tilted bipolar sunspots, involves randomness because the convective buffeting on rising flux tubes causes a scatter in the tilt angles. From the evidence present here it would appear that the standard Babcock-Leighton paradigm is sufficient to explain the connection between hemispherically asymmetric solar activity and the observed differences in times of polar field reversals in corresponding hemispheres.  In every cycle since the polar fields were first observed, the reversals have been at different times, and simply related to the prevailing activity asymmetry readily produced by a dynamo, e.g. as suggested by \cite{par71}.

\acknowledgments

We are grateful to Roger Ulrich for the use of the supersynoptic charts from MWO.


\clearpage

\clearpage


\begin{thebibliography}{}
\bibitem[Altrock(2003)]{alt03} Altrock, R. C., 2003, \solphys, 216, 343
\bibitem[Altrock(2011)]{alt11} Altrock, R. C., 2011, \solphys, 274, 251
\bibitem[Altrock(2012)]{alt12} Altrock, R. C., 2012, BAAS, 44(4), 123.03
\bibitem[Babcock(1953)]{bab53} Babcock, H. W., 1953, \apj, 118, 387
\bibitem[Babcock(1959)]{bab59} Babcock, H. D., 1959, \apj, 130, 364
\bibitem[Babcock $\&$ Babcock(1955)]{bab55} Babcock, H. W., \& Babcock, H. D., 1955, \apj, 121, 349
\bibitem[Babcock(1961)]{bab61} Babcock, H. W., 1961, \apj, 133, 572
\bibitem[Choudhuri et al.(2007)]{cho07} Choudhuri, A. R., Chatterjee, P., \& Jiang, J., 2007, \prl, 98, 131103
\bibitem[Gopalswamy et al.(2003)]{gop03} Gopalswamy, N., et al., 2003, \apj, 586, 562
\bibitem[Gopalswamy et al.(2012)]{gop12} Gopalswamy, N., et al., 2012, \apjl, 750, L42
\bibitem[Hoeksema(2012)]{hoe12} Hoeksema, J. T., 2012, BAAS, 44(4), 206.07
\bibitem[Howard(1972)]{how72} Howard, R., 1972, \solphys, 25, 5
\bibitem[Kirk et al.(2009)]{kir09} Kirk, M. S., et al., 2009, \solphys, 257, 99
\bibitem[Parker(1971)]{par71} Parker, E. N., 1971, \apj, 164, 491
\bibitem[Shiota et al.(2012)]{shi12} Shiota, D., et al., 2012, \apj, 753, 157
\bibitem[Svalgaard et al.(1978)]{sva78} Svalgaard, L., Duvall, T. L., \& Scherrer, P. H., 1978, \solphys, 58, 225
\bibitem[Temmer et al.(2006)]{tem06} Temmer, M., et al., 2006, \aap, 447, 735
\bibitem[Ulrich et al.(2002)]{ulr02} Ulrich, R. K., et al., 2002, \apjs, 139, 259
\bibitem[Waldmeier(1973)]{wal73} Waldmeier, M., 1973, \solphys, 28, 389
\bibitem[Waldmeier(1978)]{wal78} Waldmeier, M., 1978, Astron. Mitt. Eidgen. Sternw. Z\"urich, Nr. 360, 27p
\end{thebibliography}
\end{document}